\documentclass[doublespacing]{elsart}
\usepackage{natbib}
\usepackage{amssymb}
\usepackage{graphicx}
\usepackage{subeqn}

\begin{document}

\begin{frontmatter}

\title{Density-Dependence as a Size-Independent Regulatory Mechanism}
\author{Harold P. de Vladar}
\address{Theoretical Biology Group\\
Centre for Ecological and Evolutionary Studies\\
University of Groningen\\
Kerklaan 30. 9751 NN Haren\\
The Netherlands}
\ead{H.Vladar@biol.rug.nl}
\begin{abstract}

The growth function of populations is central in biomathematics. The main dogma is the existence of density dependence mechanisms, which can be modelled with distinct functional forms that depend on the size of the population. One important class of regulatory functions is the $\theta$-logistic, which generalises the logistic equation.  Using this model as a motivation, this paper introduces a simple dynamical reformulation that generalises many growth functions. The reformulation consists of two equations, one for population size, and one for the growth rate. Furthermore, the model shows that although population is density-dependent, the dynamics of the growth rate does not depend either on population size, nor on the carrying capacity. Actually, the growth equation is uncoupled from the population size equation, and the model has only two parameters, a Malthusian parameter $\rho$ and a competition coefficient $\theta$. Distinct sign combinations of these parameters reproduce not only the family of $\theta$-logistics, but also the van Bertalanffy, Gompertz and Potential Growth equations, among other possibilities. It is also shown that, except for two critical points, there is a general size-scaling relation that includes those appearing in the most important allometric theories, including the recently proposed Metabolic Theory of Ecology. With this model, several issues of general interest are discussed such as the growth of animal population, extinctions, cell growth and allometry, and the effect of environment over a population.

\end{abstract}

\end{frontmatter}
\bibliographystyle{elsart-harv.bst}

\section{Introduction}
The logistic equation is a paradigm for population biology. This simple model, in its continuous \citep{Verhulst1838,Pearl27} or discrete \citep{May76} versions describe two fundamental properties of population biology, which are (i) the initial exponential rates of growth, and (ii) density-dependent effects, like competition under limited resources, indicated by saturation values. The discrete logistic equation, in itself opened a new and broad field in biology related to chaotic behaviours, and for which some experimental evidences exist \citep{Hanski93,Gonzalez03}. The continuous version of logistic growth, although sharing properties with its discrete analog, differs in some aspects. It does not show intrinsic bifurcations as the discrete version does, and is much more simple to treat analytically.

\citet{Gilpin73} and \citet{Gilpin76} introduced a model that ``slightly'' generalises the popular logistic equation. Their model, consists on modifying the term corresponding to the density-dependence with an exponent \(\theta\): 
\begin{equation}
\dot{x} = \rho x \left[1- \left(\frac{x}{x_\infty}\right)^\theta\right]~,
\end{equation}
where $x$ is population size, the notation \(\dot{x}\) corresponds to time derivative \(dx/dt\), and \( \rho\), \(\theta\), and $x_\infty$  are parameters of the model.
 Compared to the logistic equation, their ``global model'' describes a population that converges in time to the same size as the logistic growth, i.e. to the carrying capacity $x_\infty$. However, the exponent $\theta$  gives new interpretations to this sigmoid  model of growth. If \(\theta > 1\) then intra-specific competition is high, and the population takes more time to reach its asymptotic value, termed carrying capacity. If  \(0 < \theta < 1\) then competition is lower and the carrying capacity is reached earlier than in the corresponding logistic dynamics \citep{Gilpin73,Gilpin76}.

The $\theta$-logistic model, as it has been termed afterwards, introduced a new concept on population ecology that is the $\theta$-selection strategies\citep{Gilpin73,Gilpin76}. Originally, they proposed the model to explain data from competing \emph{Drosophila} systems after failure to use a Lotka-Volterra-like model \citep{Ayala73}. Afterwards, non-competitive versions of the system (i.e. one ``allele'' or one ``species'' model) have been used  in conservation ecology to model avian population dynamics and calculate extinction times \citep{Saether00}, and also to estimate the effects of environmental stochasticity on population growth \citep{Saether02a}. Other population models have included stochasticity to aid parameter estimation and study the effect of environmental changes in caprine populations \citep{Saether02b}.
This model has also been employed in community ecology to estimate species abundance \citep{Diserud00}.  The $\theta$-logistic equation is a ``slightly more complicated model [that] yields significantly more accurate results'', using the original words of \citet{Gilpin73}.

There are, however, other kinds of regulation terms that have been successfully employed to model other kinds of populations and growth. Sigmoid curves in particular are attractive for biologists, but are not necessarily described by $\theta$-logistic equations. The \citet{VonBertalanffy66} equation, for example, is a sigmoid curve that is frequently used in allometric modelling, as well as the recently proposed (and controversial) curve derived from bioenergetic considerations by \citet{West02}. Another kind of sigmoid is given by the Gompertz equation \citep{Gompertz1825}, which was originally formulated to model human demographic data. The Gompertz equation has become an important tool in modelling tumour growth \citep{Norton76}, although applications include a wider range of topics. All of these sigmoid share the property of reaching carrying capacity, although they have different functional forms (Table \ref{table:growths}), which confere distinct dynamical properties: inflection points, critical behaviours near $x=0$, or rate of convergence to equilibrium.

However, not all populations obey saturated growth. Among non-saturated growths the first classical example is exponential growth, typically employed to describe bacterial cloning  \citep{Hershey39}, or simply as descriptors for non-regulated conditions of growth. A ``general version'' of the exponential is potential growth -which actually shows some kind regulation but does not reach a carrying capacity. Potential growth appears in tumour biology \citep{Hart98}, early-life evolution \citep{Szathmary87}, life history theory \citep{Calder84,Roff86,Day97,Stearns92}, as well as in allometry \citep{Peters83,Calder84,Brown00}. Potential growth functions is typically a consequence of complex systems where there are several levels of organisation having a direct consequence on growth (e.g. \citep{Szathmary87,West97}).

Motivated by the $\theta$-logistic equation, this paper introduces an alternative way to interpret and formulate population dynamics models. The description explained through out this paper reduces \emph{exactly} to most common population models, including the above-mentioned growth dynamics (resumed in Table \ref{table:growths}). With this new formalism general scaling laws are derived, using initial population size and carrying capacities. These scaling laws, include the heuristic scaling introduced by \citet{West02} in allometry.

%
%

\section{\label{sect:mechanics}``Mechanics'' of Self Regulation}
One of the central issues in population dynamics is to determine the growth function that describes a particular population. Growth dynamics in general can be expressed in the form
\begin{equation}
\label{eq:PopGeneralDynamics}
\dot{x} = x r(x)~.
\end{equation}

The growth rate $r(x)$ is an explicit function of the size of the population $x$. Depending on the nature of the self-regulation, $r(x)$ has different functional forms (see table \ref{table:growths}.) For a wide review of density dependence functions, including (mainly) discrete dynamics, the reader can refer to \citet{Henle04}.

\begin{table}[t]
\caption{Common regulation functions for different population growth models. In all these equations $x$ is the size of the population, $\rho$ is the Malthusian parameter, $\theta$ is the competition coefficient, and $\alpha$ is a parameter determined from environmental conditions. When populations grow to a saturation, $\alpha$ is related to the carrying capacity.}
\label{table:growths}
\begin{tabular}{lr@{=}lccc}\hline
  Model         & \multicolumn{2}{c}{\parbox{0.9in}{Growth\\ Rate}}  & \parbox{0.8in}{Malthusian Parameter} & \parbox{0.8in}{Interaction Parameter} & \parbox{0.5in}{Initial Rate} \\
\hline
 Exponential     & $r(x)$&     $\alpha$  & $\rho = 0$ & $\theta = 0$ & $\alpha(0) = \alpha \neq 0$ \\
Logistic        & $r(x)$&     \(\rho(1-\alpha x)\) & $\rho > 0$ & $\theta = 1$ & $\alpha(0) > 0$ \\
$\theta$-Logistic& $r(x)$&    \(\frac{\rho}{\theta}(1-\alpha x^{\theta})\)& $\rho > 0$ & $\theta > 0$ & $\alpha(0) > 0$\\
Gompertzian     & $r(x)$&     \(-\rho\log(\alpha x)\) & $\rho > 0$ & $\theta = 0$ & $\alpha(0) > 0$\\
Potential       & $r(x)$&     \(\alpha x^{\theta}\) & $\rho = 0$ & $\theta \neq 0$ & $\alpha(0) \neq 0$\\
von Bertalanffy & $r(x)$&  \(-3\rho(1-\alpha x^{-1/3}) \)     & $\rho >0 $ & $\theta = -\frac{1}{3}$ & $\alpha(0) > 0$\\
\citet{West02}  & $r(x)$&  \(-4\rho(1-\alpha x^{-1/4})\)    & $\rho >0 $ & $\theta = -\frac{1}{4}$ & $\alpha(0) > 0$\\
\hline
\end{tabular}

\end{table}

Growth of a population is a contribution of two terms: a replication term, whose contribution to the rate is typically a positive constant indicating the number of offspring per unit of time, or frequency of duplication, and another term describing interaction and/or growth inhibition, which is in itself dependent on $x$.

The logistic model, for example, describes growth inhibition with a second order term, i.e. $x^2$. The \(\theta\)-logistic generalises this second order term to one of an arbitrary order greater than one, expressed by \(x^{\theta+1}\). In a biological sense, this non-linear term is proportional to the frequency of interactions that an individual must have in order to produce population growth inhibition (anergy) or promotion (synergy). For these examples, the rate will be a decreasing line with $x$ for the logistic case, and a decreasing curve of order $\theta$ for the $\theta$-logistic.

The idea in this paper is to express the dynamics of a population using not only its size $x$ as the variable of interest but a decomposition where the rate $r$ at which the population grows is considered as a separate state variable. The decomposition requires the knowledge of how $r$ changes in time, thus the starting point is to determine the dynamics of $r$. This decomposition results in a two dimensional dynamical system \citep{Ermentrout02,Arrowsmith90} with the variables \((x,r) \in \mathbb{R}^+\times\mathbb{R}\) describing the population size and replication velocity, respectively.

\subsection{Exponential Growth}
In the case of the exponential growth, because $r=$ const $=\alpha$, the time derivative of the rate \(\dot{r} = 0\). Thus the following trivial system:
\begin{subequations}
\label{syst:exponential}
\begin{equation}
\label{eq:ExpGrowth}
\dot x  = x r~,
\end{equation}
\begin{equation}
\label{eq:ExpRate}
\dot r  = 0 ~,
\end{equation}
\end{subequations}
is equivalent to exponential growth.

Integrating Eq. (\ref{eq:ExpRate}) gives the constant $\alpha$, which is determined by the initial conditions of the system \(\left(x(0),r(0)\right)\). Later, substitution into Eq.(\ref{eq:ExpGrowth})  recovers the original expression in terms of one variable, i.e. \(\dot x  = x \alpha\).

\subsection{\label{sec:logistic}Logistic Growth}
For the logistic growth, it is necessary to \emph{define} the new variable $r$ as
\begin{equation}
\label{eq:implicitlogisticrate}
r(x) := \rho (1-\alpha x)~,
\end{equation}
where $\rho$ is the Malthusian parameter \citep{Fisher30}), and $\alpha>0$ is the inverse of the carrying capacity. Thus the rate equation again is expressed implicitly as \(\dot x = xr\), and the time derivative for $r$ is:
\begin{displaymath}
\dot{r} = -\rho \alpha \dot x = -\rho \alpha x r~,
\end{displaymath}
 Regrouping, and then 
summing and subtracting $1$ in the parenthesis, it is possible to write:
\begin{displaymath}
\dot{r} = \rho (1-\alpha x-1) r = \left(\rho (1-\alpha x)-\rho\right) r~.
\end{displaymath}

The inner parenthesis of the last expression has the \emph{explicit} form of $r$. After replacing it with Eq. (\ref{eq:implicitlogisticrate}), the rate equation becomes:
\begin{equation}
\label{eq:RateLogistic}
\dot{r} = (r-\rho) r~.
\end{equation}

Therefore, to solve the dynamical system equivalent to the logistic equation only one parameter and an initial condition are needed. Actually, the initial condition automatically defines the carrying capacity of the population.

\subsection{\(\theta\)-Logistic Growth.}
The rate for the \(\theta\)-logistic is defined as
\begin{equation}
\label{eq:RateTetaLogistic}
r(x) := \frac{\rho}{\theta} (1-\alpha x^{\theta})~.
\end{equation}

Following the same methodology as with the logistic growth, it is not difficult to demonstrate that the the implicit form for the rate equation is:
\begin{equation}
\label{eq:THEEquation}
\dot{r} = (\theta r -\rho) r ~.
\end{equation}

Although derived from the $\theta$-logistic, this last equation is very general because it includes all the growth laws in table \ref{table:growths}. In the limit \(\theta \rightarrow 1\) the logistic equation is recovered, and taking \emph{jointly} the limits \(\theta,\rho \rightarrow 0\) Eq. (\ref{eq:THEEquation}) reduces to the simple form of exponential growth. Note that from the logistic equation (\ref{eq:RateLogistic}) it is not possible to  take the limit to the exponential formally, since it does not show an explicit dependence on \(\theta~(= 1)\). To recover exponential growth from the explicit form of the logistic, the limit would have to be taken as \(\alpha \rightarrow 0\). But then the rate of the exponential growth will be $\rho$ instead of $\alpha$. In the formalism presented in this paper \(\alpha\) and \(\rho\) have distinct properties. On the one hand, $\rho$ is defined as a parameter of the system, and as such may have a role in bifurcations and global stability, while $\alpha$ is defined from the initial conditions, so it does not play any role in local or global stability \citep{Arrowsmith90}. Also a particular population grows following a predefined replication constant $\rho$, which is considered to be determined at least in part by intrinsic factors, while $\alpha$ is determined extrinsically by environmental conditions which define the carrying capacity of the system \citep{MacArthur62}. Thus in this mechanistic interpretation where $r$ determines growth response, the environmental issues play no dynamical role \emph{unless they are explicitly and dynamically affecting growth rate}.

%
%
\section{Generalised Growth Rates}
The three versions of the model studied above, namely exponential, logistic and $\theta$-logistic, conform just a part (in fact a minority) of the possible outcomes of the system. They were derived by some non negative combinations of parameters \(\theta \hbox{ and } \rho\). These, and other dynamics admitting also negative values for \(\theta \hbox{ and } \rho\), conform a dynamical system that generalises most classic types of population growth (Table \ref{table:growths}). In other words, the model presented herein is a unification of several growth dynamics. Resuming, population growth can be described in a general form by the two equations:
\begin{subequations}
\label{syst:global}
\begin{equation}
\label{eq:growth}
\dot x  = x r~,
\end{equation}
\begin{equation}
\label{eq:rate}
\dot r  = (\theta r -\rho)r ~,
\end{equation}
\end{subequations}
referred to from now on as \emph{growth equation} and \emph{rate equation}, respectively.

Although the rate equation (\ref{eq:rate}) indicates that regulation mechanisms are independent of population size, \emph{per capita} response is a contribution of both, individual reproduction (related to the parameter \(\rho\)) and interaction with other individuals (related to the parameter \(\theta\)).

The units of the parameter \(\rho\) are inverse of time (frequency), and it gives the characteristic time scale at which individuals down-regulates the reproduction rate when, for example, the population approaches an equilibrium state like carrying capacity or extinction. The parameter \(\theta\), is non-dimensional, but it sets the density scale at which the interaction of an individual with the population affects its reproduction rate.

%
%

\section{\label{sect:stability} Fixed Points and Stability Analysis}
The rate equation (\ref{eq:rate}) encloses all the information of the fixed points of the population dynamics. The rate equation has two fixed points, namely, \(r_0 = 0\) and \(r_1 =\rho/\theta\). Intuitively \(r_0\) corresponds to the non-trivial equilibrium point of the growth equation, i.e. when the rates become zero the population is in a stationary  equilibrium between reproduction and mortality. This means that $r_0$ is a steady state under balanced regulation.

Take for example the explicit form of the rate for the \(\theta-\)logistic equation
\[r_0 = 0 = r(x^*) = \frac{\rho}{\theta} \left(1-\alpha (x^*)^\theta\right)\] that implies \(x^*=\alpha^{-1/\theta}\), and which corresponds to the carrying capacity. In this case, the population has a finite size, balanced by reproduction (replication at the individual level) and mortality (competition at the population level).

The biological meaning of the second equilibrium point, \(r = r_1\), is not so obvious. From well-known cases, like the \(\theta-\)logistic equation, it is possible to realise that the population has a fixed point in \(x^*=0\). However, the fixed point given by the rate equation (\ref{eq:RateTetaLogistic}) means that \[r_1 = \frac{\rho}{\theta} = r(x^*) = \frac{\rho}{\theta} \left(1-\alpha (x^*)^\theta \right)\] which implies directly \(x^* = 0\). Thus \(r^*=r_1\) is equivalent to \(x^* = 0\)

The system (\ref{syst:global}) suggest a third fixed point: \((x^*,r^*) = (0,0)\). This point, however is a paradoxical point, since both, the rate and the population size cannot be simultaneously zero, unless \(\rho, \theta, \hbox{ and } \alpha\) are zero.

The stability of these fixed points, can be studied with the eigenvalues method \citep{Ermentrout02,Arrowsmith90}.
The Jacobian matrix of the system (\ref{syst:global}) is
\begin{equation}
\mathbb{J}(x,r) = \left(
\begin{array}{lcc}
r       &~~~~&       x       \\
0       &~~~~&       2\theta r -\rho
\end{array}
\right)
\end{equation}

Each of the two eigenvalues of this matrix, \(\lambda_x\) and \(\lambda_r\) are derived from evaluating the Jacobian in the fixed points \citep{Caswell00}. For the first fixed point, \(P_0=(x_0,r_0) = (x^*,0) \) (where \(x^*\) is the asymptotic value obtained by equating \(x^* = r^{-1}(0)\), like for example, carrying capacity), leads to the eigenvalues:
\begin{equation}
\label{eq:eigenzero}
\left\{
\begin{array}{ccl}
\lambda_{x_0} &=& 0\\
\lambda_{r_0} &=& -\rho
\end{array}
\right.
\end{equation}

Now, evaluation of the Jacobian matrix in the second fixed point, i.e. \hbox{\(P_1 = (x_1,r_1) = (0,\rho/\theta)\)} gives the eigenvalues:
\begin{equation}
\label{eq:eigenone}
\left\{
\begin{array}{ccl}
\lambda_{x_1} &=& \rho/\theta\\
\lambda_{r_1} &=& \rho
\end{array}
\right.
\end{equation}

A fixed points will be stable if both of its eigenvalues are negative, and unstable if any of them eigenvalues is positive. Since \(\rho\) and \(\theta\) can take any real value, the stability of the fixed points \(P_0\) and \(P_1\) depends only the signs of these two parameters, but not in their magnitudes.

However, some properties are already evident. First, none of the fixed points can be foci, since the eigenvalues cannot take imaginary values, thus oscillations and cycles cannot occur. Second, the fixed point $P_0$ has a null eigenvalue, \(\lambda_{x_0} = 0\) (Eq. \ref{eq:eigenzero}), indicating that there is a invariant set \(x^{inv}\) that has ``null" stability, meaning that the point at which every trajectory intersects \(x^{inv}\), (a) depends entirely on the initial conditions \(\left(x(0),r(0)\right)\), and (b) is fixed. This implies that  distinct orbits intersect \(x^{inv}\) at different points. The dynamical equations imply that \(x^{inv}\)(also termed stable manifold) corresponds to \(r = 0\) for all non-negative values of size $x$.

\begin{figure}
\includegraphics[scale=0.7]{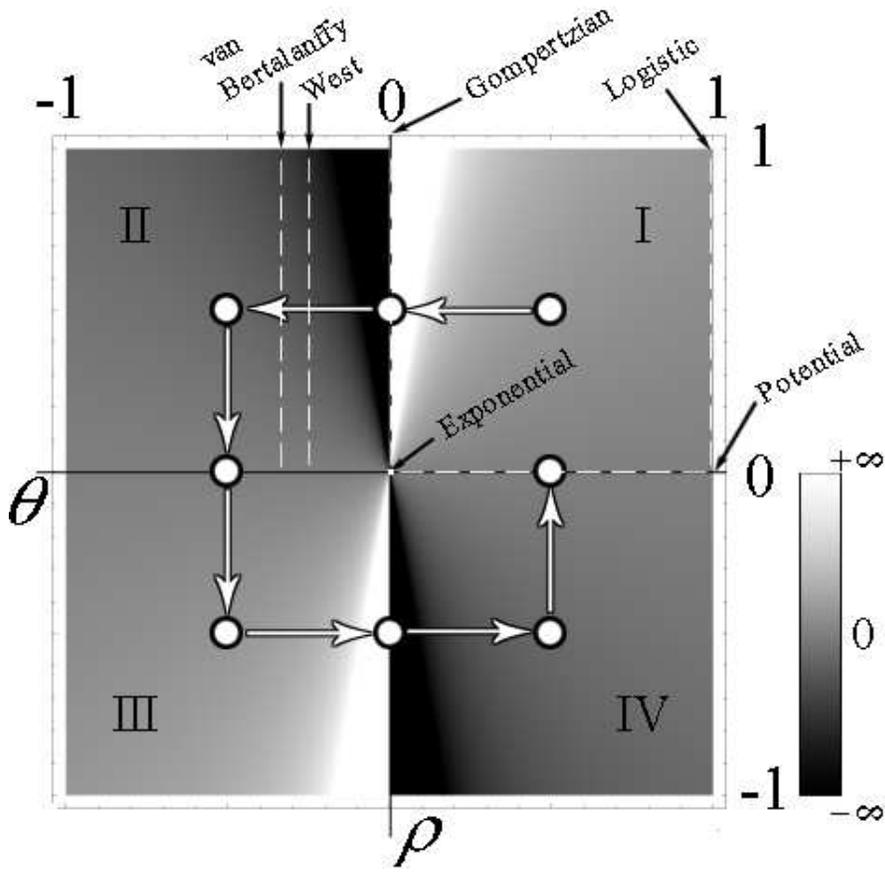}
\caption{The value of the fixed point $r_1$ (shown as a shade) as a function of the parameters \(\rho\) and \(\theta\). The four dots inside the four quadrants correspond to the main four classes of models, and the four dots on top of the axes are bifurcation points. The arrows follow the explanation in section \ref{sect:stability}. In quadrant I, $r_1$ is an unstable node; the point at \(\theta=0\), following the arrow, is a bifurcation point ($r_1$ is at infinity). In quadrant II, $r_1$ is a saddle. Following the arrow to \(\rho=0\) another bifurcation point is found. In this point, the invariant set \(x^{inv}\) changes its stability from stable (attractive) to unstable. In quadrant III, $r_1$ continues to be a saddle. Following the arrow  to the point at \(\theta=0\) another bifurcation is found. The stability of $r_1$ changes in quadrant IV to a stable node. Finally, returning to quadrant I, another bifurcation is found at \(\rho=0\) where $r_1$ changes to an unstable node, and \(x^{inv}\) becomes stable again. The white dashed lines represent the growth rates listed in table \ref{table:growths}.}
\label{fig:Rfixed}
\end{figure}

According to the signs of \(\rho\) and \(\theta\), Eqs. (\ref{eq:eigenzero}-\ref{eq:eigenone}) indicate that there are four distinct possible sign combinations for the eigenvalues. For each of these combinations, termed regimes, particular patterns in the trajectories occur. This suggests several types of equilibria, convergence to equilibrium (which corresponds to growth dynamics), and transitions between the distinct regimes or types of equilibria (bifurcations), as shown in figure \ref{fig:Rfixed}.

\begin{figure}[t]
\includegraphics[scale=0.4,angle=-90]{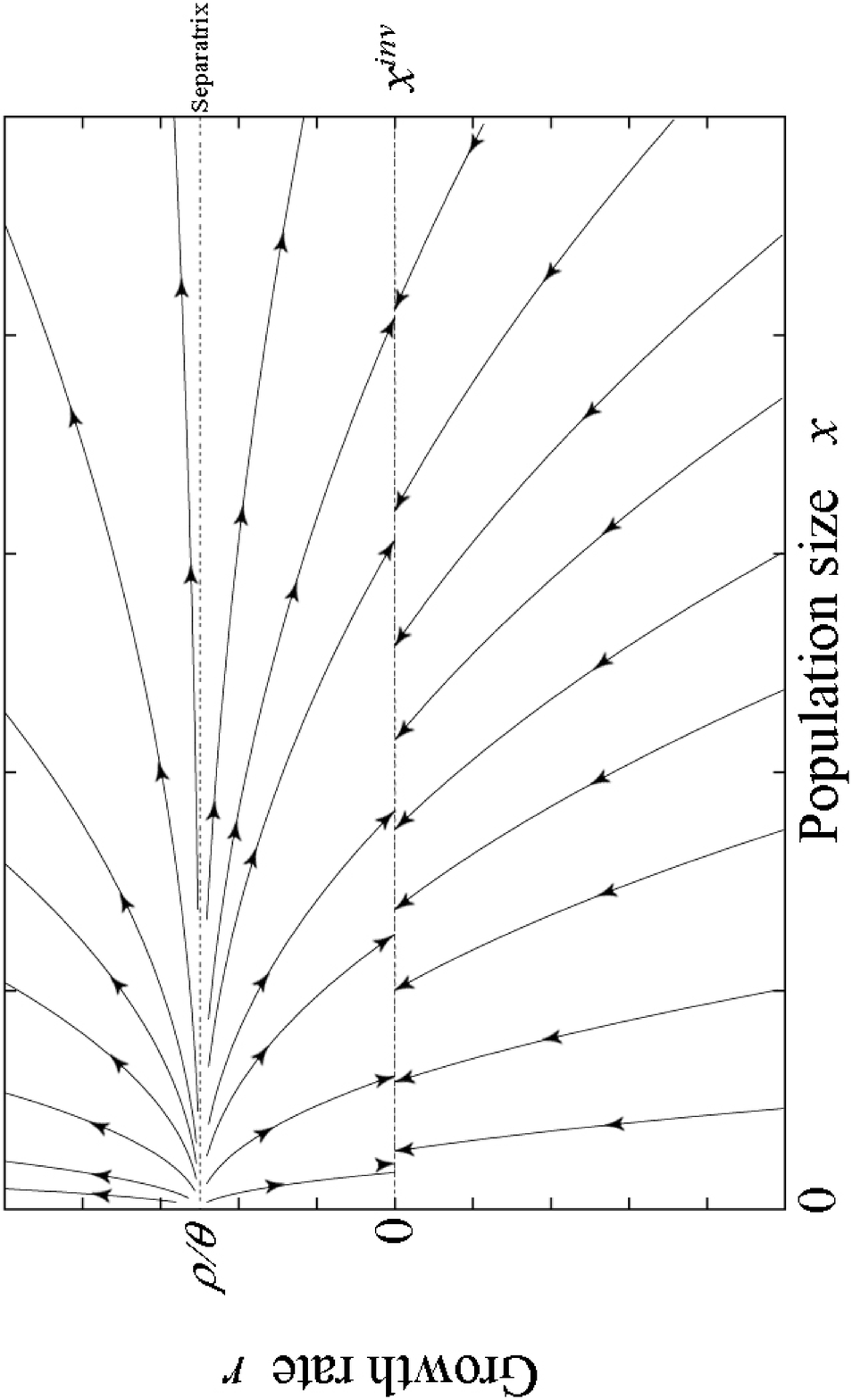}
\caption{This regime of the phase space corresponds to the first quadrant in figure \ref{fig:Rfixed}, where \(\rho,\theta>0\). There are two types of growth. If the initial conditions are below the separatrix  \(r(0) < \rho/\theta\), then the population grows as a $\theta$-logistic. The intersection of the orbits at $r=0$ correspond to carrying capacities. Initial conditions the are above the separatrix (\(r(0) > \rho/\theta\)), growth in a cooperative way, where interactions between individuals accelerate population increase.}
\label{fig:regimen1}
\end{figure}

 %
 %
The first quadrant in figure \ref{fig:Rfixed} where \(\rho\) and \(\theta\) are both positive, corresponds to a phase space that has an unstable node at $P_1 = \rho/\theta$. The trajectories that start below this point, i.e. \(r(0) < P_1\) are attracted to \(x^{inv}\) (Fig. \ref{fig:regimen1}.) However, if the initial conditions are \(r(0) > P_1\), then the orbits are upper unbounded, and growth is unlimited. This means that the line \(r = \rho/\theta\) is the separatrix for the two possible dynamics.

Maintaining \(\rho > 0\) and decreasing \(\theta \rightarrow 0\), the systems shows a discontinuous (i.e. first-order) transition (Fig. \ref{fig:Rfixed}). Because the position of the fixed point \(P_1 = \rho/\theta\) increases as  $\theta$ decreases, at \(\theta = 0\) this $P_1$ disappears at infinity. In this case, when $\theta = 0$, \(x^{inv}\) retains its stability (note that the eigenvalues associated to \(x^{inv}\), \(\lambda_{x_0}\) does not depend on \(\theta\)), and all orbits in the system converge to \(x^{inv}\), for all initial conditions.

 %
 %

\begin{figure}[t]
\includegraphics[scale=0.4,angle=-90]{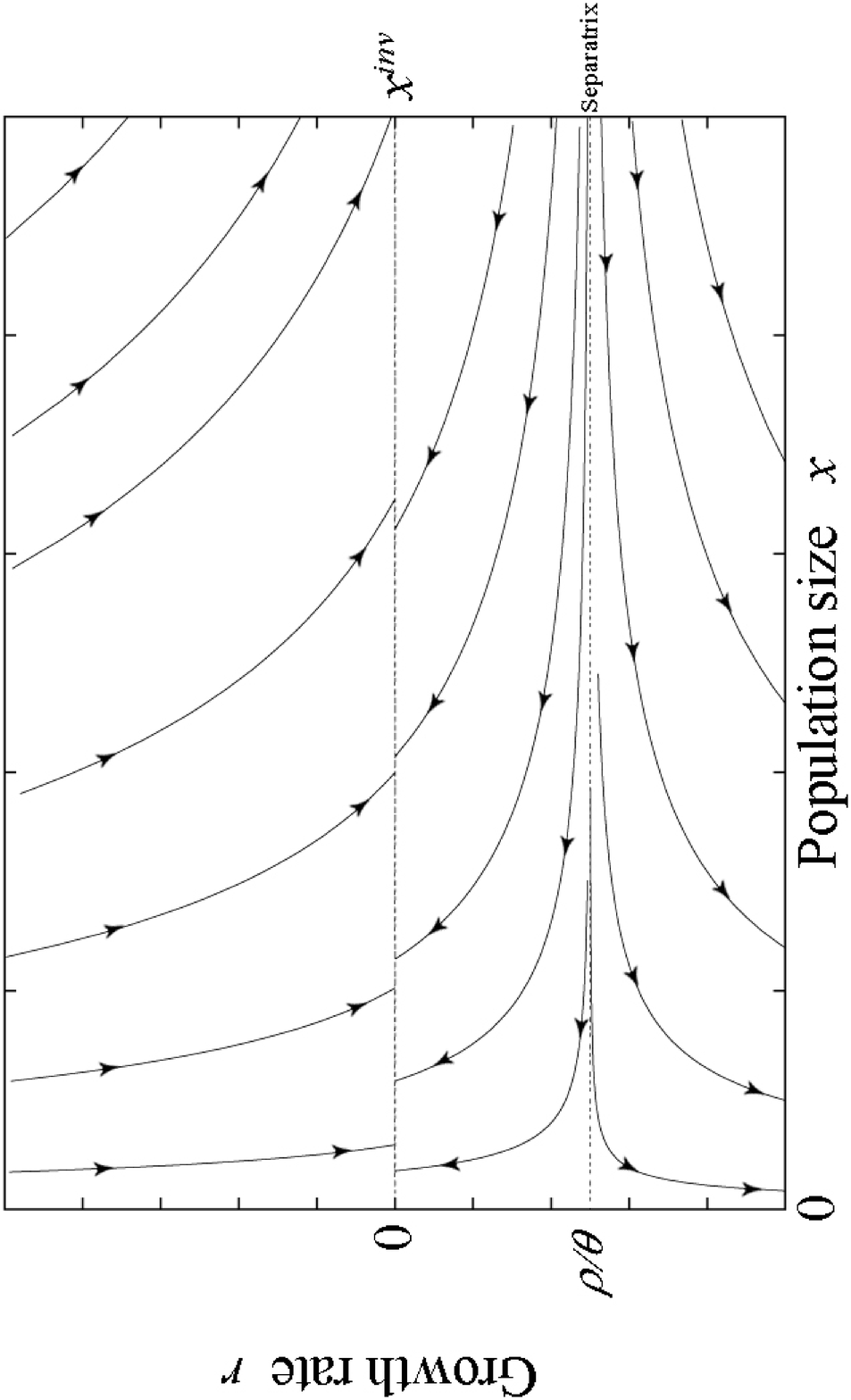}
\caption{In this regime, corresponding to quadrant II in figure \ref{fig:Rfixed} with $\rho>0$ and  $\theta<0$ two types of growth are possible. The initial conditions above the separatrix at $r=\rho/\theta (<0)$ reproduce sigmoid growth curves which converge to a carrying capacity that corresponds to the intersection of the orbits at $r=0$. If the initial conditions are below the separatrix, then the interactions are anergistic and the population decreases hyperbolically , and become  extinct in a finite time.}
\label{fig:regimen2}
\end{figure}

When \(\theta < 0 \) -and maintaining \(\rho > 0\)- the system is in quadrant II (Fig. \ref{fig:Rfixed}), and its properties change:  \(P_1\) is a saddle (i.e. unstable) point (now there is one positive eigenvalue in Eq. \ref{eq:eigenone}.) The invariant manifold \(x^{inv}\) still retains it attracting stability (Fig. \ref{fig:regimen2}).

If the initial conditions  \(r(0)> - \rho/\theta\), then the orbits intersect \(x^{inv}\) as in the previous regime (Fig. \ref{fig:regimen2}). Note that \(\theta < 0\), thus these initial conditions include a region where  negative rates diverge  (meaning then that population decreases in size toward zero).

From this point in the second quadrant of the parameter space (Fig. \ref{fig:Rfixed}), decreasing \(\rho\) to zero while maintaining  \(\theta<0\) leads to a continuous (or second order) transition. Here the stable varieties \(x^{inv}\) and the separatrix \(P_1\) collapse onto each other.

Because at this point \(\lambda_{x_1} = \lambda_{r_1} = 0\), we cannot infer about the dynamical properties of the nullcline at $r=0$ with the eigenvalues method. However, with a perturbation on each side of the fixed point, it is possible to determine the stability. Then if \(\Delta r\) is the perturbation (a trajectory slightly displaced from the fixed point), the rate equation for the regime where \(\rho = 0\) and \(\theta < 0\) is
\begin{displaymath}
\dot{\Delta r} = -|\theta| \Delta r^2~,
\end{displaymath}
thus the system always responds diminishing the rate. Consider the solution $R(t)$ of the perturbation to the rate equation (note that there is no ``first order" approximation in the rate equation for this case):
\begin{displaymath}
R(t) = \frac{\Delta r}{1+\Delta r |\theta| t}~.
\end{displaymath}

If the perturbation is positive, the rate will be damped to zero asymptotically. This means that the population will grow potentially. If the perturbation is negative, the rate will decrease to \(-\infty\), in a finite time given by \(t_e = (\Delta r |\theta|)^{-1}\).

Thus \(x^{inv}\) repels the orbits on its left (initial conditions \(r(0) < 0\), and asymptotically attracts the orbits on its right (initial conditions \(r(0) > 0\)).

 %
 %

\begin{figure}
\includegraphics[scale=0.4,angle=-90]{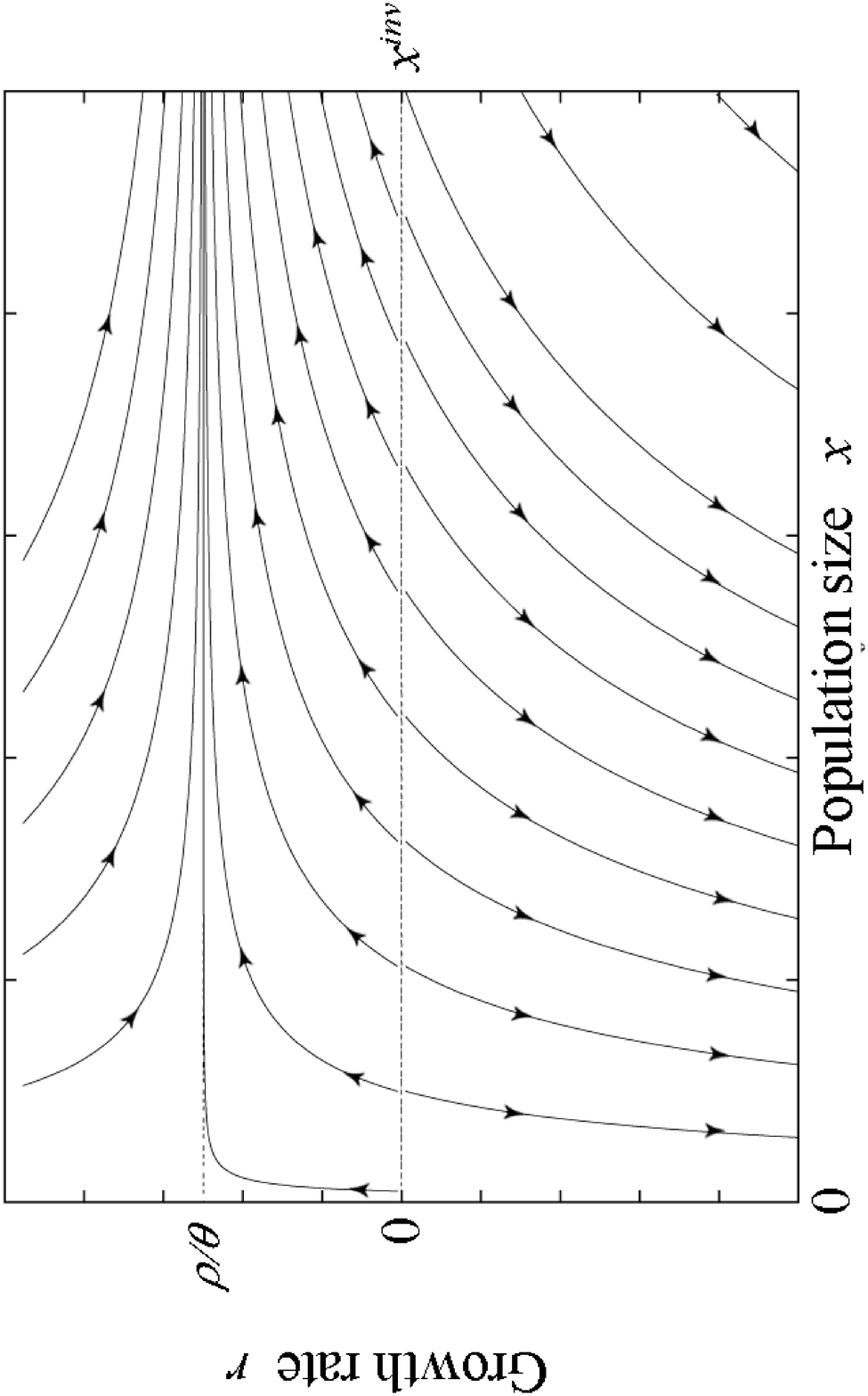}
\caption{An invasion-extinction regime, corresponding to quadrant III of figure \ref{fig:Rfixed}, where $\rho,\theta<0$. This regime represents dynamics that show threshold behaviour. If the initial rate is $r(0) > 0$ the population can invade asymptotically-exponential with rate $\rho/\theta$. If initial conditions are \(r(0) < 0\) the population may go extinct in finite time with the rate decreasing hyperbolically.}
\label{fig:regimen3}
\end{figure}

In the third quadrant of the parameter space, where \(\theta < 0\) and \(\rho < 0\), the eigenvalue \(\lambda_{x_1} > 0\) but \(\lambda_{r_1}<0\) (according to the relationships Eqs. \ref{eq:eigenone}). Thus $P_1$ is again a saddle point. In this case the trajectories grow \(x\rightarrow \infty\), and \(x^{inv}\) is a separatrix that repels the orbits on its neighbourhoods. Figure \ref{fig:regimen3} shows several trajectories for this regime.

Proceeding in direction to the fourth quadrant, maintaining \(\rho < 0 \) and decreasing \(\theta \rightarrow 0\), once again a discontinuous transition is found, where \((x,r) \rightarrow \infty\), and \(x^{inv}\) still repels the trajectories from each side. 

 %
 %

\begin{figure}
\includegraphics[scale=0.4,angle=-90]{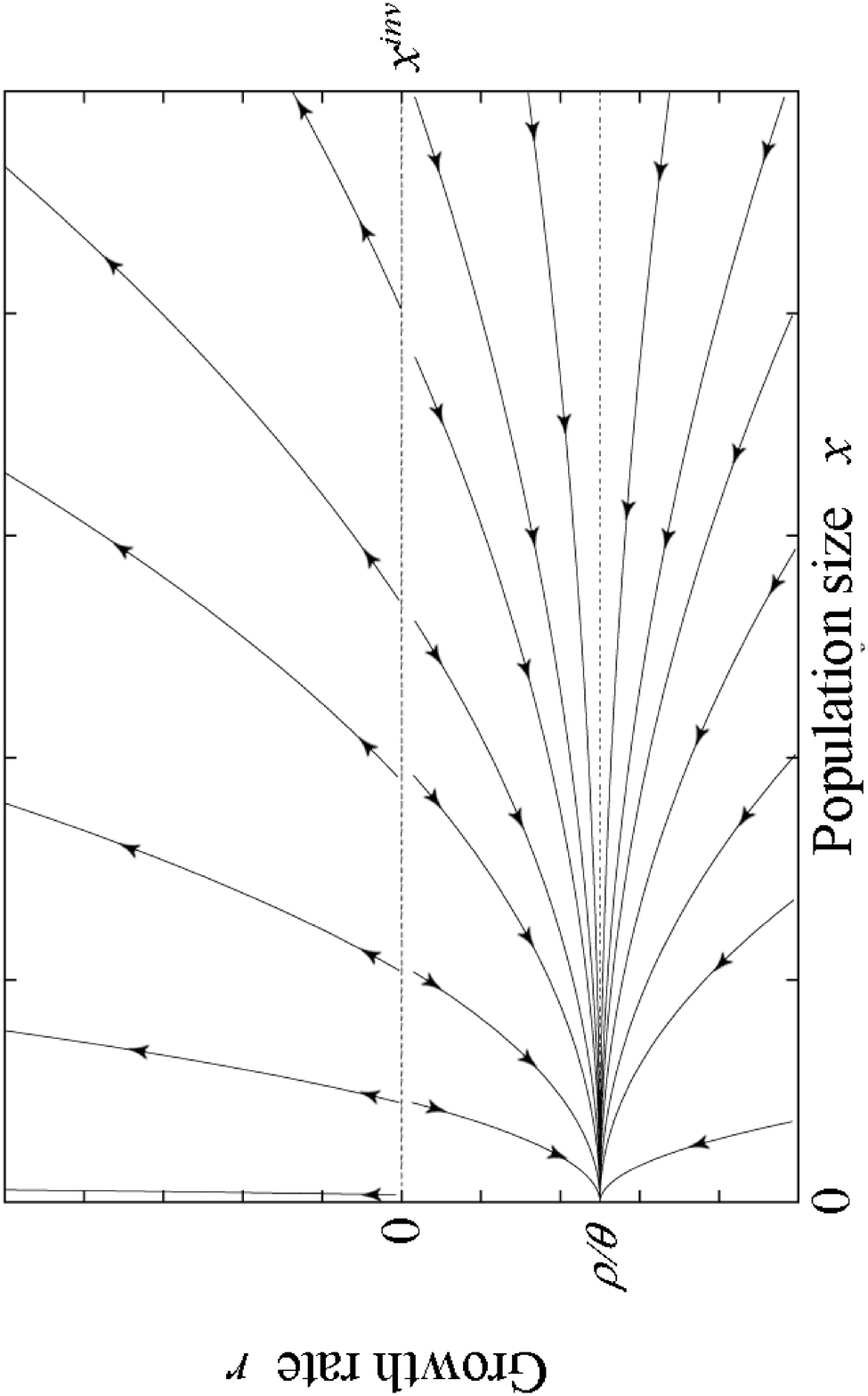}
\caption{An invasion-extinction regime corresponding to quadrant IV in figure \ref{fig:Rfixed} with \(\rho<0,\theta>0\). This regime is of threshold type. Invasion occurs in a synergistic way when \(r(0) > 0\), although when the conditions are met for extinction, i.e. \(r(0)<0\) the population vanishes decreasing exponentially.}
\label{fig:regimen4}
\end{figure}

The last regime when \(\theta > 0\) and \(\rho < 0\), is characterised by being the only one where $P_1$ is a stable point (Fig. \ref{fig:regimen4}). \(x^{inv}\) is a separatrix: for those initial conditions such that \(r(0) < 0\) (i.e. negative rates), the populations decrease to zero sigmoidally, and if the initial conditions are \(r(0) > 0\), then the population increases hyperbolically to \(x\rightarrow\infty\). 

%
%

\section{Carrying Capacity}
In the growth dynamics, the explicit rate equations (Table \ref{table:growths}) involve a constant \(\alpha\) which does not appear in the implicit form of the dynamical system (\ref{syst:global}). Integrating the growth equation leads naturally to the constant $\alpha$, as it did in the calculations of the exponential and  $\theta$-logistic equations in section \ref{sect:mechanics}. Once the initial conditions are defined the constant \(\alpha\) is determined. In quadrants I and II \(x^{inv}\) is stable, and the point at which the orbits intersect \(x^{inv}\) correspond to the carrying capacity of the system,  \(x_\infty\). The value of \(\alpha\) is related to the carrying capacity as: \(x_\infty = (\alpha)^{-1/\theta}\). Moreover, all sigmoid dynamics are represented in these two quadrants. When \(\theta >0\) the trajectories correspond to logistic and $\theta$-logistic dynamics, whose applications range from populations of flies \citep{Gilpin73} to mammals \citep{Saether02b}. For $\theta = 0$ the dynamics correspond to Gompertzian growth, which has been widely applied in tumour biology to investigate distinct aspects of tumour response and regression \citep{Norton76,Norton77}, as well as microbiological models \citep{Kozusko03}. Whenever $\theta < 0 \) ontogenetic growth laws, like the model by \citet{VonBertalanffy57}, or the model by \citet{West02}, are included in this regime.

Thus by itself $\alpha$ plays no particular role in the stability of the system. In this formalism the growth rate is an adaptive mechanism that responds to the environmental conditions, which enter as the initial conditions. Different initial conditions result in different growth models. Previous formulations included the carrying capacity as a predefined constant, entering explicitly in the model as the parameter $\alpha$ or $x_\infty$. This constrains the dynamics to a particular model, and also suggests that the growth of a population is not dependent of the environment \citep{MacArthur62}. In this formalism the growth rate is an adaptive mechanism that responds to the environmental conditions. The same model can grow toward distinct carrying capacities (or even change its qualitative form to an unlimited growth) just by changing the initial conditions. Explicit models with predefined carrying capacity, need extra assumptions for model selection. In this formulation the mechanism for regulation remains robust against environmental changes.

\section{\label{sec:scaling}Scaling Laws}

Several systems that look very different may actually have exactly the same dynamics when they are properly scaled in time and size. Scaling is common phenomenon in physics \citep{Kadanoff}. However, in biology only recently scaling hypotheses in dynamical processes are being proposed \citep{West97,Brown00,West02,Brown04,Kozlowski04}. Whenever systems can be scaled, it is no longer important (in the dynamical sense) the precise values of the constants. The growth law of the population remains invariant whenever proper relationships between the constants and the variables are considered. 

To determine the scaling laws for the solutions of the system (\ref{syst:global}) it is convenient to replace \(r\rightarrow \theta r\), which gives
\begin{subequations}
\begin{equation}
\dot{x} = \frac{1}{\theta} r x~,
\end{equation}
\begin{equation}
\dot{r} = r(r-\rho)~.
\end{equation}
\end{subequations}

Denoting the solution for the system (\ref{syst:global}) as \((X(t),R(t))\) growth equation yields
\begin{equation}
\label{eq:gensol}
X(t) := x(0) \exp\left(\frac{1}{\theta} \int_0^t R(s) ds \right)~,
\end{equation}
where \(x(0)\) is the initial condition for $x(t)$. Rearranging this system, we get
\begin{equation}
\left(\frac{X(t)}{x(0)}\right)^\theta = \exp\left( \int_0^t R(s) ds \right)~.
\end{equation}

The right-hand side of the last equation is independent of \(\theta\). Thus populations described by the system (\ref{syst:global}) are always scalable to their initial sizes, and interaction exponent $\theta$ (except, maybe for the bifurcation points.)

Further rescaling is possible for the right-hand side. The solution to the rate equation can be written in the form
\begin{equation}
\label{eq:ratesolution}
R(t) := \rho \left[1+ e^{\rho t}\left(\frac{\rho}{r(0)}-1\right)\right]^{-1}~,
\end{equation}
and changing the time variable as
\begin{equation}
\label{eq:magictransform}
T \rightarrow -\rho t - \log \left(\frac{\rho}{r(0)}-1\right) ~,
\end{equation}
and also changing  properly the differential in the integral in Eq. (\ref{eq:gensol}) to \(dT' = \rho dt\), then the result of the integral is, in scales of \(T\)
\begin{equation}
\label{eq:protolaw}
\left(\frac{X(t)}{x(0)}\right)^\theta =  \frac{1+e^{T_0}}{1+e^T}~.
\end{equation}

The variable $T$ depends on the initial condition \(r(0)\) whose meaning is not so obvious.

Although scaling laws are general, in practice (e.g. in allometry) scaling laws are relevant when populations reach a carrying capacity. Suppose then that the population achieves a carrying capacity \(x_\infty\) given by \(x_\infty = \alpha^{-1/\theta}\). Using the explicit form of \(r(t)\) (Eq. \ref{eq:RateTetaLogistic}), it results that the term inside the logarithm in the rescaled time (Eq. \ref{eq:magictransform}) is:
\begin{equation}
\label{eq:importantrelation}
\left(\frac{\rho}{r(0)}-1\right)^{-1} = \left(\frac{x(0)}{x_\infty}\right)^{-\theta} - 1~.
\end{equation}

Substituting the relation (\ref{eq:importantrelation}) into the time transformation  (\ref{eq:magictransform}), and rearranging terms in equation (\ref{eq:protolaw}):
\begin{equation}
\left(\frac{x(t)}{x_\infty}\right)^{-\theta} = 1 -  \left[ \left(\frac{x(0)}{x_\infty}\right)^{-\theta} - 1\right] e^{-\rho t} ~.
\end{equation}

To conclude, define the new scaled variables as:
\begin{equation}
\label{eq:ScaledVariables}
\left\{
\begin{array}{ccl}
\chi &=& \left(\frac{x(t)}{x_\infty}\right)^{-\theta} \\
\tau &=& \rho t - \log \left[ 1- \left(\frac{x(0)}{x_\infty}\right)^{-\theta} \right] \\
\end{array}
\right.
\end{equation}
with which the general scaling law obeys
\begin{equation}
\label{eq:THELAW}
\chi = 1- e^{-\tau}~.
\end{equation}

This scaled dynamical expression is valid for all (non zero) sign combinations of $\rho$ and $\theta$.
 %
 %

\subsection{Scaling at the bifurcation points}
The scaling law (Eq. s\ref{eq:THELAW}) cannot be defined for Gompertzian or potential growth, because the general solutions for the size and rate equations (\ref{syst:global}) are not necessarily equations (\ref{eq:gensol}) and (\ref{eq:ratesolution}). It is necessary to proceed calculating the solutions of Gompertzian and potential growth, and then derive the scaling laws.

\subsubsection{Potential Growth} The solution for the potential growth dynamics is given also by Eq. (\ref{eq:gensol}). Consider then, that the solution to its rate equation is
\begin{equation}
R(t) := r(0) \left( 1 - r(0) t \right)^{-1}~.
\end{equation}

In this case, the initial condition $r(0)$ is not expressed in terms of a carrying capacity.  Potential growth or decay does not reach an asymptotic value. Thus  $r(0)$  can only be defined in terms of the integration constant $\alpha$. Upon integration and rearranging of terms, the following form is found:
\begin{equation}
\left(\frac{X(t)}{x_0}\right)^{-\theta} = 1 - \alpha t ~.
\end{equation}

The scaled variables can then be defined as
\begin{equation}
\left\{
\begin{array}{ccl}
\chi &=& \left(\frac{X(t)}{x_0}\right)^{-\theta} \\
\tau &=& \alpha t \\
\end{array}
\right.~,
\end{equation}
with which the general scaling law obeys
\begin{equation}
\label{eq:THELAWscaledPot}
\chi = 1-\tau~,
\end{equation}
that is simply a decreasing line.

\subsubsection{Gompertzian Growth} Consider now the solution to the size equation for the Gompertzian Growth
\begin{equation}
\frac{X(t)}{x(0)} = \exp \int_0^t R(t') dt'~.
\end{equation}

The main difference between Eq. (\ref{eq:gensol}) and the last equation is that the former can be scaled with the exponent \(\theta\), which for the Gompertzian Growth case is zero.

Consider then the solution \(R(t)\) to the rate equation:
\begin{equation}
R(t) := r(0) e^{-\rho t} ~.
\end{equation}

Thus the general solution for Gompertzian Growth is
\begin{equation}
\label{eq:GompertzGenSol}
\frac{X(t)}{x(0)} =  \exp \left[ \log\left( \frac{x(0)}{x_\infty} \right) \left(e^{-\rho t} - 1\right) \right]~,
\end{equation}
where the initial condition $r(0)$is expressed  as:
\begin{equation}
r(0) = -\rho \log\frac{x(0)}{x_\infty}~.
\end{equation}

Eq. (\ref{eq:GompertzGenSol}) can be rearranged to give
\begin{equation}
\log\frac{x(t)}{x_\infty} = \exp\left[-\rho t + \log \log \frac{x_\infty}{x(0)}\right]~.
\end{equation}

Thus, defining the dimensionless variables:
\begin{equation}
\left\{
\begin{array}{ccl}
\chi &=& \log(x_\infty/x(t)) \\
\tau &=& \rho t - \log \log (x_\infty/x(0)) \\
\end{array}
\right.
\end{equation}
then the scaled dynamics results as
\begin{equation}
\label{eq:THELAWscaledGomp}
\chi = e^{-\tau}~,
\end{equation}
that is a decreasing exponential.

Although the scaling behaviour is completely defined for all the dynamics in the phase space, it is still  interesting to determine how the scaling law Eq. (\ref{eq:THELAW}) is related to the scaling laws Eqs. (\ref{eq:THELAWscaledPot},\ref{eq:THELAWscaledGomp}).

Direct evaluation of \(\rho = 0\) or \(\theta = 0\) does not give the scaling laws for Gompertzian or potential growth. However, Taylor expansion \emph{on the parameters} $\rho$ and $\theta$ to the linear term around the critical values, result in the scaling laws of the critical points.

Note that in order to give a precise meaning to the transformations Eqs. (\ref{eq:ScaledVariables}), the initial condition for the rate equation was transformed to its explicit form. It is necessary, however, to make this transformation \emph{after} the limits are taken in the scaled variables Eqs. s(\ref{eq:ScaledVariables}).

%
%

\section{Discussion}
The simple model derived in this paper is rich in qualitative solutions since it resumes several growth rates that often appear in the literature,  which include several levels of biological organisation. The examples alluded in the text range from cellular populations of procariots, cellular populations in eucariots, in ontogeny and cancer, to population biology of mammals and birds, to community ecology. It is a nice result that all of these kinds of growth can be described by such simple equations that resumes the main features of populations, in the traditional sense of showing density dependence, and in the distinct interpretations introduced in this paper.


\subsection{Extinctions and Invasions}

Environmental changes, are know to ``unbalance'' some populations. In this model, environment enters as initial conditions; translations of the $r$-component in the phase space can model changes in the environment. Suppose that a population is in its carrying capacity, and suddenly the trajectory is perturbed. This will placed the trajectory in an orbit out of equilibrium. Thus the population will converge to a new carrying capacity. However, if the perturbation is strong enough, then the orbit where the dynamics is placed could belong to a basin of attraction leading to extinction or invasion.

As a first example, consider global warming. The metabolic theory of ecology, proposed by \citet{Brown04} proposed considers that the carrying capacity of a population is temperature-dependent through a Boltzmann factor. This dependence can be expressed as \hbox{\(x_\infty = K_0 \exp(E/kT)\)}, where  \(K_0\) is a parameter depending on mass, resources, etc., $E$ is the energy of the limiting metabolic reaction, $k$ is Boltzmann's constant, and $T$ the absolute temperature.

Suppose a population that is in its carrying capacity at a temperature
$T_0$  \(x_\infty = K_0 \exp(E/kT_0)\). If suddenly the temperature increases to $T_1$, the change in the rate due to temperature increase can be calculated, introducing the Boltzmann factors for population size in the explicit form of rate (e.g. Eq. \ref{eq:RateTetaLogistic}):
\begin{equation}
\label{eq:westwarming}
\Delta r=\frac{\rho}{\theta}\left[1-\exp\left(\frac{\theta E}{kT_0 T_1}(T_1-T_0)\right)\right]<0~.
\end{equation}
This means that the population is taken out from equilibrium. Carrying capacities exist whenever \(\rho > 0\) (for any value of $\theta$). According to Eq. (\ref{eq:westwarming}) The population will attain a new lower carrying capacity. Thus the model of \citet{Brown04} and this formulation imply that temperature  cannot induce population extinction. To induce an extinction the exponential term would have to change sign, which is not possible for any temperature (it is always possible however, to consider such changes that although they theoretically do not imply extinctions, numerically are so small that in real life populations could disappear.)

Kin selection, is a second example. The saturated dynamics, comprised in quadrants I and II of figure \ref{fig:Rfixed}, have other co-existing behaviours which support  Hamilton's rule  . which points out that if the cost of an altruistic behaviour is such that it benefits a genetically related individuals, then the strategy can be selected \citep{Hamilton63,Hamilton64}. In this way, a population consisting of cooperative individuals can grow faster than expected by exponential models. This can happen when $\alpha<0$ in quadrant I (Fig. \ref{fig:regimen1}), because growth is accelerated as a result of the interaction between the individuals. Notice that for \(\alpha'<\rho/\theta\)
\begin{equation}
\frac{\rho}{\theta}\left(1+|\alpha|x^\theta\right) > \alpha'
\end{equation}
for all $x$; and even if \(\alpha'>\rho/\theta\), there exists some $\underline{x}$ such that for $x>\underline{x}$ the inequality holds. Thus in this region of the phase space, population growth is faster than exponential.

However, Hamilton's rule has another consequence. This is that aggressive behaviours can also be selected, provided that damage is induced to ``negatively related'' (i.e. unrelated) individuals (as it has been reported for wasps \citet{Gardner04}.) This kind of behaviour is recovered for quadrant II (Fig. \ref{fig:regimen2}) where any interaction between individuals results in mutual annihilation indicated by $\alpha > 0$s. When the initial conditions  are \(r(0) < - \rho/\theta\),  the rate decreases.  Solving the rate equation (\ref{eq:rate}),  shows that \(r \rightarrow -\infty\) asymptotically (Fig. \ref{fig:regimen2}) when \(t \rightarrow t_e\), where
\begin{displaymath}
t_e = \frac{1}{\rho} \log\left( \frac{r(x_0)}{ r(x_0)-\frac{\rho}{\theta}} \right) ~,
\end{displaymath}
thus extinction occurs in finite time.

Another example, mentioned in the text above, is the extinction of sparrows \citep{Saether00}. In order to have the risk of finite-time extinctions from previously stable populations, it is necessary that (a) the population dynamics belongs to quadrant II of figure \ref{fig:Rfixed}, and (b) there is a perturbation such as mentioned above. The estimated mean value for the population of sparrows is $\hat{\theta} \simeq 1$, indicating that the population is logistic. However, the estimate distribution for $\theta$ allows a small but not negligible probability for \(-1.5 < \theta < 0\). If this is the case, then a real risk of finite time extinction exists.

\subsection{Life histories}
In life history theory, the central problem is resource allocation for adaptive strategies. Survival and reproduction in distinct stages determine of the net growth of an individual \citep{Day97,Stearns92}. In non-reproductive stages of life, i.e. before maturity, resources are mainly devoted to growth. The energetic content allocated to growth depends on the size of the body. This dependence follows a potential growth function, where the exponent $\theta$ is indicative of some length scale of the physiological processes \citep{Calder84,Stearns92}.

According to the rate equation (\ref{eq:rate}), initial stages in development have to be dominated by the term $\theta r^2$, which implies that resources are devoted to growth. This term is dominant at low densities, when the exponent \(\theta < 0\). This condition is met in in the \citet{VonBertalanffy57} and \citet{West02} equations, and well as in most allometric growth relationships \citep{Calder84}. When reproduction becomes a priority, less energy is devoted to growth. In this case the term $\rho r$ is not negligible, indicating that the organism is partitioning the resources between growth and survival.

\subsection{Allometry and scaling}
The scaling laws derived in section \ref{sec:scaling} are general formulas showing that scaling is rather a rule than an exception. This gives an broader view to the open discussion  of whether the \citet{West02} equation is legitimate or not \citep{Kozlowski04}. In terms of these formulations, although numerically different, West's equations and its classical competitor, van Bertalanffy's equation, have the same qualitative behaviour. 

The model (\ref{syst:global}) reproduces the \citet{VonBertalanffy66} equation. This equation can be written in the following form:
\begin{equation}
\label{eq:VB}
\dot x = ax^{2/3} - bx ~.
\end{equation}
It is possible to rearrange this equation, to express it as the system (\ref{syst:global}), for which the parameters are then:
\begin{equation}
\begin{array}{ccc}
\rho = \frac{b}{3} &,& \theta = -\frac{1}{3}~.
\end{array}
\end{equation}
The exponent \(-1/3\) follows from the hypothesis that mass is proportional to the third power of length, and the parameter $b$ is related to individual reproduction. The parameter $a$ is related to the carrying capacity of the population, thus it does not appear in the transformations.

Another example in this regime is West's ontogenetic growth equation \citep{West02}, given by
\begin{equation}
\dot x = a m^{1/4} \left[1-\left(\frac{x}{x_\infty}\right)^{1/4}\right]~,
\end{equation}
 The parameters for the rate equation are then:
\begin{equation}
\begin{array}{ccc}
\rho = \frac{a}{4 x_\infty^{1/4}} &,& \theta = -\frac{1}{4}~.
\end{array}
\end{equation}
The exponent $-1/4$  derived from fractal patterns of fluid transport systems like circulatory system or plant vascularisation \citep{West97,Brown04} and is also supported by empirical data \citep{West02}. 

The parameter \(\rho\) in this case depends explicitly on the carrying capacity \(x_\infty\). This suggests that in ontogenetic growth, these two quantities could be correlated, implying that the macroscopic growth i.e. the cell population, is ``transmitting'' information to the microlevel, i.e. single-cell dynamics and thus it could imply existence of self-organisation (this has been demonstrated for Gompertzian growth \citep{Molski03}. 

In both models, van Bertalanffy and West, \(\rho > 0\) and \(\theta < 0\), belonging to the same region in the parameter space (Fig. \ref{fig:Rfixed})


\subsection{Cancer}

The growth of tumours have been very well studied through distinct kinds of mathematical and computational models \citep{Wheldon88}. \emph{In vitro} and experimental frameworks have shown that they grow according to a Gompertzian Law \citep{Norton76}. The Gompertzian law, a sigmoid curve that grows toward a carrying capacity, implies that resources in a tumour are devoted to reproduction, as indicated by $\theta = 0$. In the rate-based scheme the orbits are attracted to the invariant set \(x^{inv}\) for any initial condition.
  
  \citet{Molski03} demonstrated that the Gompertz equation is the result of self-organisation (cooperativity), in such a way that the response of each cell is coherently correlated to the state of the whole population. In their model, \citet{Molski03} found a Gompertzian regression rate (shrinking) solution. In the formalism presented inhere the Gompertzian regression solution corresponds to $\rho<0$ and $\theta<0$ with initial conditions $r(0) < 0$. This kind of decreasing size tumours appear under external perturbations (which actually leave the rate equation invariant) are applied, like radiation or chemotherapy treatments \citep{Gonzalez03B,PdeVladar04}.



\section{Concluding Remarks}
The $\theta$-logistic equation has become a paradigm in ecology. Modelling populations with it has been an important tool to confront actual problems about density-dependent ecology. The transformation introduced in this paper gives a good insight into the meaning of the quantities appearing in the equations, namely $\rho$, $\theta$, and $\alpha$ (either in its interpretations as carrying capacity or not). 

There are, of course, more examples for each of the growth types described in this paper. However, more interesting is that there are regimes that have not been reported. This is not surprising, because they conform distinct types of indeterminate growth, which usually is assumed to be ``exponential''. However, these distinct types of explosions, can have important consequences in disciplines like biotechnology, where a strict control of growth is necessary. If for some reason, a population is wrongly manipulated, such that it spreads ``indeterminately'' then the distinct types of growths should be managed distinctly.

However, in the opinion of the author, the most important result is that the rate equation is explicitly independent of the population size. The results presented in this paper, are derived from a simple mathematical transformation,  which surprisingly results in a very broad class of regulatory mechanism. Although this is a result that may apply only to the simple systems included in this work, it is puzzling why and how the regulatory mechanisms act.

The two terms of the rate equation (\ref{eq:rate}), from a more abstract perspective, correspond to two processes that constitute regulation: reproduction, which comes from an individual level, and sensitivity to population interaction. Depending on the context, the sensitivity to the population can be of synergistic or anergistic nature. The two regulation processes, could be thought of as fragmentation and condensation reactions:
\begin{equation}
\begin{array}{cccl}
x  &\smash{ \mathop{\longrightarrow}\limits^\rho}&  2 x &\hbox{(reproduction)}\\
N x &\longrightarrow&  M x & \hbox{(population sensitivity)}
\end{array}
\end{equation}
If $M>N$ then the population experiences synergy in growth (i.e. population interaction promotes growth). If $M<N$ then the population experiences anergy in growth (i.e. population interactions avoid growth). 

The term $-\rho r$ in the rate equation indicates that the population is growing or ``relaxing'' to a fixed point. Since $\rho$ is the inverse of the relaxation time for the rate, then the bigger $\rho$ is the smaller the time to let the mechanism to relax, and thus the fastest to reach limiting population size at $x^{inv}$. 

The second term of the rate equation is related to the interaction between individuals in the population. This term can be compared to a ``potential'' indicating some kind of resource potentiating (either synergistically or anergistically) from the interaction. The kind of interaction, is given by the sign of $\theta$ and by the environmental conditions, i.e. by $\alpha$. 

The relationship between $\rho$ and $\theta$ determining the distinct types of growth rates, gives distinct types of behaviours for distinct initial conditions which can result in cooperative, competitive, or aggressive strategies. These are strategies that can be sought directly from the rate equation. If a population is behaving cooperatively, then it means that the rate begins over a threshold such that it grows unlimited, because there is a benefit improving growth resulting from the interaction. In this case, the resources have to be unlimited, so cooperativity improves resource allocation for reproduction.  In the case of competition, usually the scenario is that resources are limited, and there must be an equilibrium between reproduction and survival. But if the population presents aggressive behaviours, then the initial rate of the population is below the threshold where it goes extinct in finite time or exponentially.
  

The exponential growth is a particular case in which no regulation mechanism (thus interactions) is present in the population. Increase is based only in individual reproduction, and the \emph{per capita} response is totally independent of the state of the system.

The distinct qualitative solutions (regimes) for population growth have an underlying symmetry. The regression equation for population shrinking is in general obtained by inverting the time arrow, changing $t \rightarrow -t'$. However, time-inverting the rate equation does not produce the desired result. In order to obtain the regression dynamics from the size-rate decomposition, besides inverting the time arrow, it is necessary to invert the rate variable $r\rightarrow -r'$. Thus a time reversed equation results in the transformed system

\begin{subequations}
\label{syst:globalinverted}
\begin{equation}
\label{eq:growthinverted}
\dot x  = x r'~,
\end{equation}
\begin{equation}
\label{eq:rateinverted}
\dot r'  = (\theta r' +\rho)r' ~.
\end{equation}
\end{subequations}

Notice that these transformations are equivalent to change the sign of the Malthusian parameter $\rho$. This transformation for population shrinking, can be regarded as a reflexion of the parameter space (Fig. \ref{fig:Rfixed}) with respect to the $\theta$-axis. Thus, although most of the relevant population dynamics are in quadrants I and II in figure \ref{fig:Rfixed}, their corresponding regression dynamics are in quadrants III and IV.

The decomposition presented in this work, is a change in the paradigm of population dynamics. The Eqs. (\ref{syst:global}) are very general, but still simplistic because there are many biological aspects left aside. Take for example the Allee effect \citep{Allee31}. This  is a density dependent growth mechanism, not represented in the rate equation. Actually, including Allee effect in population growth, leads to a polynomial rate equation of higher order than 2. There is however no general law that can be derived (at least, including the forms found in the literature). Other example in non-random mating, which could lead to distinct forms of density dependence, maybe through a frequency-dependent parameter $\theta$. In this case the form of the rate equation derived here will not necessary be valid. 

It remains to investigate based on life history theory for which kind of resource allocation the density-independent rate equation can be derived. This is a work currently under development that is expected to help to give conclusions about other biologically relevant aspects not included in this work.

\paragraph*{Acknowledgements}
The author would like to thank Dr. I. Pen and Dr. M. Hinsch for important discussions and remarks in the original manuscript, and to Dr. D. Drasdo for illustrating discussions about tumour replication rates, held in the frame of the Thematic Institute and Workshop on Multi-Agent Systems, Max-Plank-Institut f\"ur Physik komplexer Systeme - Dresden, September 2004.


\end{document}